\documentstyle[11pt,epsfig]{article}
\begin{document}
\title{Multiplicity fluctuation analysis of target residues in nucleus-emulsion collisions at a few hundred MeV/nucleon}
\author{{Dong-Hai Zhang\thanks{Corresponding author. Tel: +863572051347; fax: +863572051347. E-mail address:zhangdh@dns.sxnu.edu.cn}, Yan-Ling Chen, Guo-Rong Wang, Wang-Dong Li, Qing Wang}\\
{Ji-Jie Yao, Jian-Guo Zhou, Su-Hua Zheng, Li-Ling Xu, Hui-Feng Miao, Peng Wang}\\
{Institute of Modern Physics, Shanxi Normal University, Linfen 041004, China}}

\date{}
\maketitle

\begin{center}
\begin{minipage}{145mm}
\vskip 0.4in
\begin{center}{\bf Abstract}\end{center}
{Multiplicity fluctuation of the target evaporated fragments emitted in 290 A MeV $^{12}$C-AgBr, 400 A MeV $^{12}$C-AgBr, 400 A MeV $^{20}$Ne-AgBr and 500 A MeV $^{56}$Fe-AgBr interactions is investigated using scaled factorial moment method in two-dimensional normal phase space and cumulative variable space, respectively. It is found that in normal phase space the scaled factorial moment ($ln<F_{q}>$) increases linearly with increase of the divided number of phase space ($ln{M}$) for lower {\em q}-value and increases linearly with the increase of $ln{M}$ and then becomes saturated or decreased for higher {\em q}-value, in cumulative variable space $ln<F_{q}>$ decreases linearly with increase of $ln{M}$, which indicates that no evidence of non-statistical multiplicity fluctuation is observed in our data sets. So any fluctuation indicated in the results of normal variable space analysis is totally caused by non-uniformity of single-particle density distribution.}\\

{\bf Keywords:} heavy-ion collisions, target fragmentation, non-statistical fluctuation, nuclear emulsion\\
{\bf PACS:} 25.70-z, 25.70.Mn, 24.60.Kg, 29.40.Rg
\\
\end{minipage}
\end{center}

\vskip 0.4in
\baselineskip 0.2in

\section{Introduction}

Since Bialas and Peschanski\cite{bialas} proposed the method of the scaled factorial moment (SFM) to study non-statistical fluctuation in multiparticle production, a lot of experiments were performed to search for the power law behavior not only for relativistic produced particles (mostly pions) but also for target fragments (target evaporated fragments and target recoiled protons) in nucleus-nucleus collisions at high energies.

Intermittency is a manifestation of scale invariance of the physical process and randomness of the underling scaling law, which is defined as the power law growth of SFM with decreasing phase-space interval size. The unique feature of this moment is that it can detect and characterize the non-statistical density fluctuations in particle spectra, which are intimately connected with the dynamics of particle production.

In high energy nucleus-nucleus collisions, target fragmentation also may carried out the information about colliding mechanism. The non-statistical
fluctuations of the emission of target evaporated fragments in lepton-nucleus, hadron-nucleus and nucleus-nucleus collisions at high energies have been investigated not only in one-dimensional phase-space[2-11] but also in two-dimensional phase-space[12-21]. The evidence of non-statistical fluctuation of target evaporated fragments is obtained in most of these investigation, but in some of these studies\cite{ghosh1,ghosh3,zhang3} the saturation effect is observed in the dependence of the power law growth of the SFM and decreasing phase-space interval size,  these effects indicate that there is not a non-statistical fluctuation in the emission of target evaporated fragments. So the study of non-statistical fluctuation of target evaporated fragments in high energy nucleus-nucleus collisions is not conclusive. For target fragmentation at intermediate and high energy (a few hundreds MeV/nucleon) nucleus-nucleus collisions, a little attention is paid to study the evaporated fragment multiplicity distribution and multiplicity fluctuation.

In this paper the non-statistical fluctuations of target evaporated fragments produced in 290 A MeV $^{12}$C-AgBr, 400 A MeV $^{12}$C-AgBr, 400 A MeV $^{20}$Ne-AgBr and 500 A MeV $^{56}$Fe-AgBr interactions are studied in two-dimensional phase space using not only normal variables $\cos\theta$ and $\phi$ but also cumulative variables $X_{\cos\theta}$ and $X_{\phi}$, where $\theta$ is the emissional angle and $\phi$ is the azimuthal angle of target evaporated fragment in the laboratory system.

\section{ Experimental details}

Four stacks of nuclear emulsion made by Institute of Modern Physics, Shanxi Normal University, China, are used in present investigation. The emulsion stacks were exposed horizontally at HIMAC, NIRS, Japan. The beams were 290 A MeV $^{12}$C, 400 A MeV $^{12}$C, 400 A MeV $^{20}$Ne and 500 A MeV $^{56}$Fe respectively, and the flux was 3000 ions/$cm^{2}$. BA2000 and XSJ-2 microscopes with a 100$\times$ oil immersion objective and 10$\times$ ocular lenses were used to scan the plates. The tracks were picked up at a distance of 5 mm from the edge of the plates and were carefully followed until they either interacted with emulsion nuclei or escaped from the plates. Interactions which were within 30 ${\mu}m$ from the top or bottom surface of the emulsion plates were not considered for final analysis. All the primary tracks were followed back to ensure that the events chosen do not include interactions from the secondary tracks of other interactions. When they were observed to do so the corresponding events were removed from the sample.

In each interaction all of the secondaries were recorded which include shower particle, target recoiled proton (grey track particle), target evaporated fragment (black track particle) and projectile fragments. According to the emulsion terminology\cite{powell}, the particles emitted from interactions are classified as follows.

(a) Black particles ($N_{b}$). They are target fragments with ionization $I>9I_{0}$, $I_{0}$ being the minimum ionization of a single charged particles. Range of black particle in nuclear emulsion is $R<3$ mm, velocity is $v<0.3c$, and energy is $E<26$ MeV.

(b) Grey particles ($N_{g}$). They are mostly recoil protons in the kinetic energy range $26\leq{E}\leq375$ MeV and a few kaons of kinetic energies $20\leq{E}\leq198$ MeV and pions with kinetic energies $12\leq{E}\leq56$ MeV. They have ionization $1.4I_{0}\leq{I}\le9I_{0}$. Their ranges in emulsion are greater than 3 mm and have velocities within $0.3c\leq{v}\leq0.7c$.

The grey and black particles together are called heavy ionizing particles ($N_{h}$).

(c) Shower particles ($N_{s}$). They are produced single-charged relativistic particles having velocity $v\geq{0.7c}$. Most of them belong to pions
contaminated with small proportions of fast protons and K mesons. It should be mentioned that for nucleus-emulsion interactions at a few hundred MeV/nucleon most of shower particles are projectile protons not pions.

(d) The projectile fragments ($N_{f}$) are a different class of tracks with constant ionization, long range, and small emission angle.

To ensure that the targets in nuclear emulsion are silver or bromine nuclei, we have chosen only the events with at least eight heavy ionizing track particles($N_{h}\geq8$).

\section{Analysis methods}

The non-statistical multiplicity fluctuation analysis is performed in a two-dimensional phase space. The phase space is divided equally in both directions assuming that is isotropic in nature. Denote the two phase space variables as $x_{1}$ and $x_{2}$, then horizontal SFM of order {\em q} is defined as\cite{bialas}

\begin{equation}
  F_{qi}(\delta{x_{1}}\delta{x_{2}})=M^{q-1}\sum_{m=1}^{M}\frac{n_{mi}(n_{mi}-1)\cdots(n_{mi}-q+1)}{n_{i}(n_{i}-1)\cdots(n_{i}-q+1)}
\end{equation}
where $\delta{x_{1}}\delta{x_{2}}$ is the size of a two-dimensional cell, $n_{mi}$ is the multiplicity in the {\em m}th cell of the {\em i}th event, $n_{i}$ is the multiplicity of the {\em i}th event, and {\em M} is the number of two-dimensional cells into which the considered phase space has been divided. Then the averaged horizontal SFM becomes

\begin{equation}
  <F_{q}(\delta{x_{1}}\delta{x_{2}})>=\frac{1}{N_{ev}}\sum_{i=1}^{N_{ev}}F_{qi}(\delta{x_{1}}\delta{x_{2}})
\end{equation}

    Non-statistical multiplicity fluctuation would manifest itself as a power-law scaling of $<F_{q}>$ with the cell size of the form

\begin{eqnarray}
   <F_{q}(\delta{x_{1}}\delta{x_{2}})>\propto(\delta{x_{1}}\delta{x_{2}})^{-a_{q}} &        & \mbox{as    $\delta{x_{1}}\delta{x_{2}}\rightarrow0$}
\end{eqnarray}
or a linear relation
\begin{equation}
  \ln<F_{q}>=-a_{q}\ln(\delta{x_{1}}\delta{x_{2}})+b_{q}=a_{q}\ln{M}+c_{q}.
\end{equation}
    The invariant quantity of the scaling $a_{q}>0$ is called the intermittency exponent, and it is a measure of the fluctuation strength.

The single-particle density distribution in two-dimensional space (emission angle space and azimuthal angle space) is non-flat. As the shape of this distribution influences the scaling behavior of the SFMs. Bialas and Gazdzicki\cite{bialas1} introduced "cumulative" variable which drastically reduced the distortion of intermittency due to non-uniformity of single-particle density distribution. Following them, the cumulative variable $X(x)$ is related to the single-particle density distribution $\rho(x)$ through

\begin{equation}
    X(x)=\int_{x_{1}}^{x}\rho(x')dx'/\int_{x_{1}}^{x_{2}}\rho(x')dx'
\end{equation}
where $x_{1}$ and $x_{2}$ are two extreme points of the distribution $\rho(x)$. The variable $X(x)$ varies between 0.0 and 1.0, with $\rho(X(x))$ kept almost constant. The values of $x_{1}$ and $x_{2}$ are -1 and 1 in $\cos\theta$-space, 0 and $2\pi$ in azimuthal angle space, respectively.

\section{Results and discussion}

Fig. 1 and 2 show the emission angle and azimuthal angle distributions of target evaporated fragment emitted from 290 A MeV $^{12}$C-AgBr, 400 A MeV $^{12}$C-AgBr, 400 A MeV $^{20}$Ne-AgBr and 500 A MeV $^{56}$Fe-AgBr interactions respectively. It is found that the angular distributions are not uniformity in whole phase space. The emission angle distribution is fitted by a Gaussian distribution, which is plotted in smooth curve. The azimuthal angle distribution are symmetric around $\phi=180$ degree.

Fig. 3 shows the dependence of $\ln<F_{q}>$ on $\ln{M}$ for target evaporated fragments emitted from 290 A MeV $^{12}$C-AgBr, 400 A MeV $^{12}$C-AgBr, 400 A MeV $^{20}$Ne-AgBr and 500 A MeV $^{56}$Fe-AgBr interactions in normal variable space. It can be seen that for lower order {\em q} the $\ln<F_{q}>$ increase linearly with $\ln{M}$ increase, but for higher order {\em q} the $\ln<F_{q}>$ increase linearly with increase of $\ln{M}$ and then saturated or decreased. So from the results of Fig.3 we can not get a clear evidence of non-statistical multiplicity fluctuation.

Fig. 4 shows the dependence of $\ln<F_{q}>$ on $\ln{M}$ for target evaporated fragments emitted from 290 A MeV $^{12}$C-AgBr, 400 A MeV $^{12}$C-AgBr, 400 A MeV $^{20}$Ne-AgBr and 500 A MeV $^{56}$Fe-AgBr interactions in cumulative variable space.  It can be seen that $\ln<F_{q}>$ decrease linearly with $\ln{M}$ increase, which indicate that no evidence of non-statistical multiplicity fluctuation is observed in our data sets. So any fluctuation indicated in Fig. 3 is totally caused by non-uniformity of single-particle density distribution.

In intermediate and high nucleus-nucleus collisions target fragmentation produces "grey" and "black" tracks in nuclear emulsion. The grey tracks are formed due to fast target protons of energy ranging up to 400 MeV. The black tracks are images of target evaporated particles of low-energy ($E<30 MeV$) singly or multiply charged fragments. In the cascade evaporation model\cite{powell}, the grey tracks are emitted from the nucleus very soon after the instant of impact, leaving the hot residual nucleus in an excited state. Emission of black particles from this state takes place relatively slowly. In the rest system of target nucleus, the directions of the emission of evaporation particles are distributed isotropically. Non-statistical multiplicity fluctuation is not existed in the state of statistical equilibrium. This is the reason we have observed in Fig. 3 and Fig. 4.

\section{Conclusions}

The emission angle and azimuthal angle distributions of target evaporated fragment emitted from 290 A MeV $^{12}$C-AgBr, 400 A MeV $^{12}$C-AgBr, 400 A MeV $^{20}$Ne-AgBr and 500 A MeV $^{56}$Fe-AgBr interactions are investigated respectively. It is found that the angular distributions are not uniformity in whole phase space. The non-statistical multiplicity fluctuation of target evaporated fragments produced in 290 A MeV $^{12}$C-AgBr, 400 A MeV $^{12}$C-AgBr, 400 A MeV $^{20}$Ne-AgBr and 500 A MeV $^{56}$Fe-AgBr interactions are studied not only in normal variable space but also in cumulative variable space, no evidence of non-statistical fluctuation in emission of target evaporated fragments is found in our data sets.

\section*{Acknowledgment}
We are grateful to Dr. N. Yasuda and staff of HIMAC, Japan for helping to expose nuclear emulsion. This work has been supported by the Chinese National Science Foundation under Grant No. 11075100, the Natural Science Foundation of Shanxi Province, China under Grant No. 2011011001-2, and the Shanxi Provincial Foundation for Returned Overseas Chinese Scholars, China under Grant No. 2011-058.

\newpage
Caption of figures:\\

Fig.1, Emissional angle distribution of target residues produced in 290 A MeV $^{12}$C-AgBr, 400 A MeV $^{12}$C-AgBr, 400 A MeV $^{20}$Ne-AgBr and 500 A MeV $^{56}$Fe-AgBr interactions. Each distribution is fitted by a Gaussian distribution, which is plotted in smooth curve.\\

Fig.2, Azimuthal angle distribution of target residues produced in 290 A MeV $^{12}$C-AgBr, 400 A MeV $^{12}$C-AgBr, 400 A MeV $^{20}$Ne-AgBr and 500 A MeV $^{56}$Fe-AgBr interactions.\\

Fig.3, Plots of $\ln<F_{q}>$ against $\ln{M}$ for the target residues produced in 290 A MeV $^{12}$C-AgBr, 400 A MeV $^{12}$C-AgBr, 400 A MeV $^{20}$Ne-AgBr and 500 A MeV $^{56}$Fe-AgBr interactions in normal two-dimensional variable space.\\

Fig.4, Plots of $\ln<F_{q}>$ against $\ln{M}$ for the target residues produced in 290 A MeV $^{12}$C-AgBr, 400 A MeV $^{12}$C-AgBr, 400 A MeV $^{20}$Ne-AgBr and 500 A MeV $^{56}$Fe-AgBr interactions in cumulative variable space.

\newpage
\begin{figure}[hp]
\includegraphics{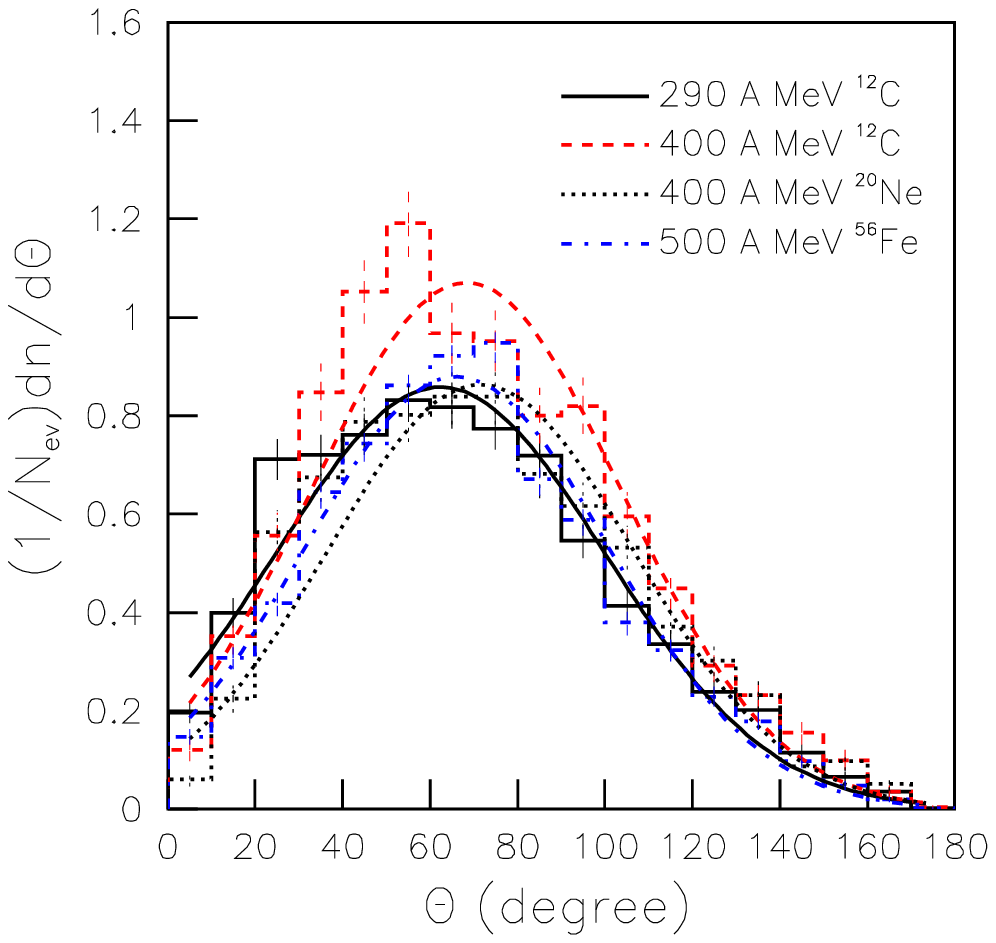}
Fig.1
\end{figure}

\begin{figure}[hp]
\includegraphics{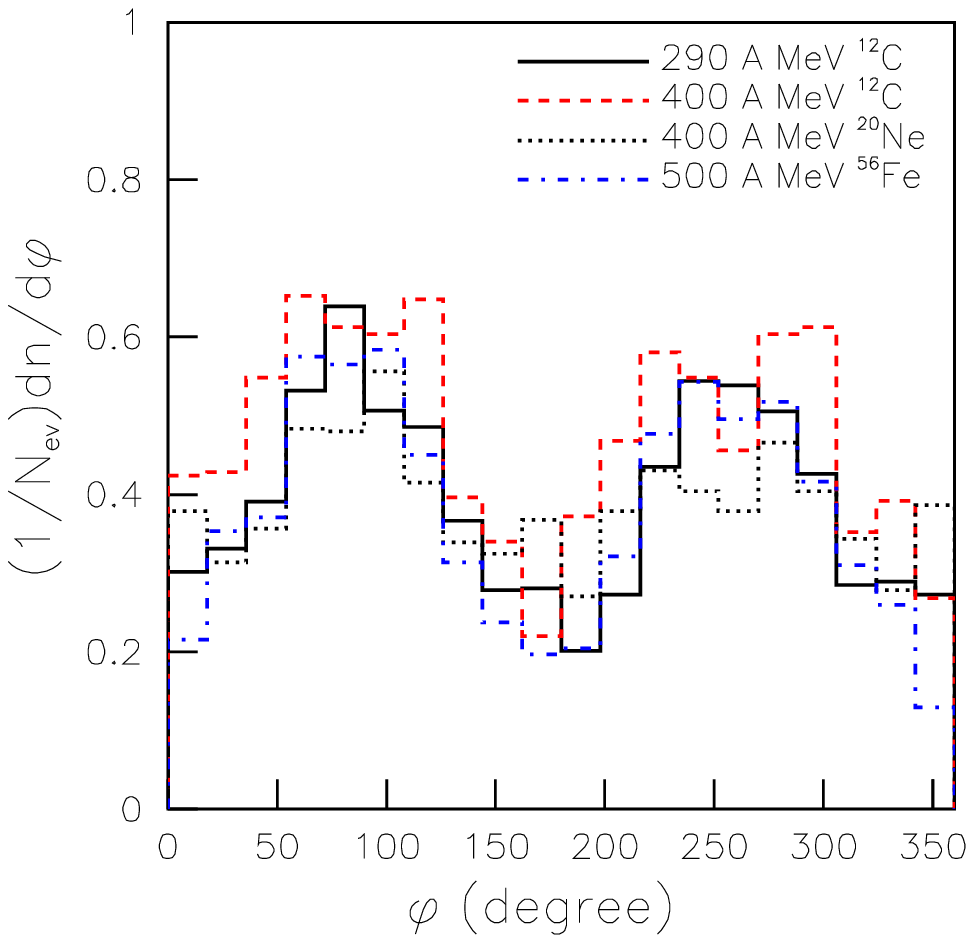}
Fig.2
\end{figure}

\begin{figure}[hp]
\includegraphics{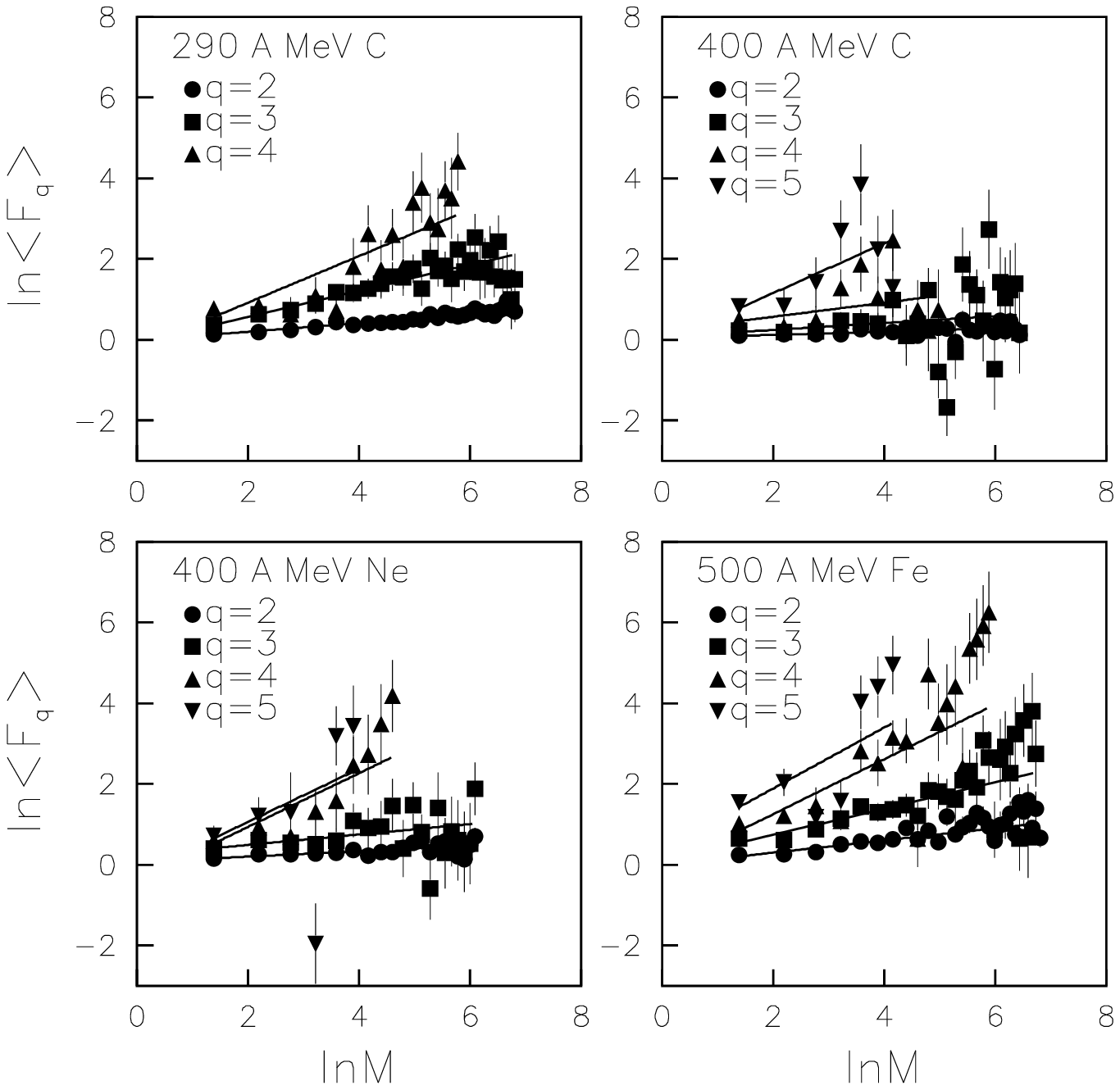}
Fig.3
\end{figure}

\begin{figure}[hp]
\includegraphics{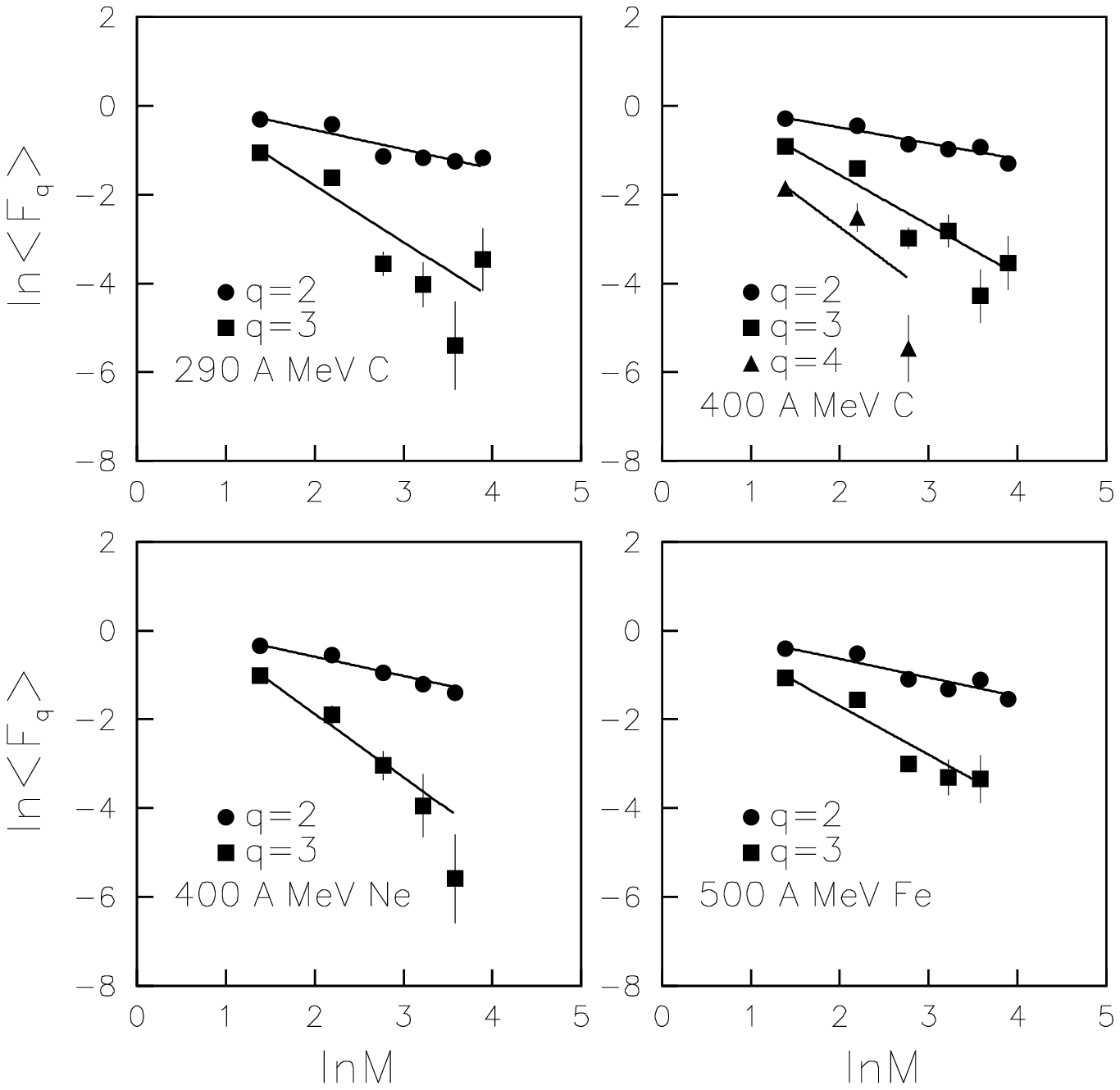}
Fig.4
\end{figure}

\end{document}